\documentclass[preprint]{revtex4}
\usepackage{graphicx}
\usepackage{bm}
\begin{document}
\setcounter{page}{1}
\title
{Extraction of the proton charge radius from experiments}
\author
{N. G. Kelkar$^1$, T. Mart$^2$ and M. Nowakowski$^1$}
\affiliation{
$^1$ Departamento de F\'isica, Universidad de los Andes,
Cra.1E No.18A-10, Santafe de Bogot\'a, Colombia\\
$^2$ Departemen Fisika, FMIPA, Universitas Indonesia, Depok 16424, Indonesia}
\begin{abstract}
Static properties of hadrons such as their radii and other moments of the electric 
and magnetic distributions  
can only be extracted using theoretical methods and not directly measured 
from experiments. As a result, discrepancies between the extracted 
values from different precision measurements can exist. 
The proton charge radius, $r_p$, which is either 
extracted from electron proton elastic scattering data or from hydrogen atom 
spectroscopy seems to be no exception. The value $r_p = 0.84087(39)$ fm extracted 
from muonic hydrogen spectroscopy is about 4\% smaller than that obtained from 
electron proton scattering or standard hydrogen spectroscopy. The resolution of 
this so called proton radius puzzle has been attempted in many different ways over the 
past six years. The present article reviews these attempts with a focus on 
the methods of extracting the radius.  
\end{abstract}
\maketitle

\section{Introduction} 
The structure of the proton plays an important role in atomic physics
where experiments have reached a very high precision. 
The inclusion of the proton structure
plays an important role in the 
accurate comparison of experimentally measured transition energies and 
very precise Quantum Electrodynamics (QED) calculations. 
Conversely, the unprecedented
precision of atomic physics experiments allows one to probe some static properties
of the proton such as its radius. 
Properties such as its charge and magnetization density are usually obtained as Fourier 
transforms of the Sachs form factors \cite{sachs} which are extracted from 
electron proton (e-p) 
scattering cross section measurements. One can deduce the radius and 
other moments from these densities and infer on the size of the proton. 
The radius thus extracted from e-p scattering and hydrogen spectroscopy seemed to 
be commensurate within error bars until a recent precision measurement of
transition energies in muonic hydrogen changed the scenario.
A comparison of the theoretical calculation of the Lamb shift in muonic hydrogen  
including all QED and finite size corrections (FSC) with the very precisely measured
value of the shift $\Delta E = E_{2P_{{3}/{2}}}^{f=2}
-E_{2S_{{1}/{2}}}^{f=1} = 206.2949(32)$ meV, in muonic hydrogen, surprisingly led to
a radius which was 4\% smaller than the average CODATA value of $0.8768(69)$ fm 
\cite{codata}. 
The extracted value of $r_p = 0.84184(67)$ fm was much more accurate than the previous 
ones. This so-called ``proton puzzle" was later reinforced \cite{antogpohl}
with the precise value of $r_p = 0.84087(39)$ fm from muonic hydrogen spectroscopy. 

The puzzle gave rise to extensive literature which attempted solutions involving 
different approaches for the evaluation of finite size corrections \cite{weNPB}, 
off-shell correction to the photon-proton vertex \cite{millerthomas}, 
the charge density being poorly constrained by data \cite{sick}, the existence of 
non-identical protons \cite{terry} as well as problems in choosing the reference frame 
in the extraction of the radius \cite{weJHEP,othersLorentz}. On the experimental side, 
accurate spectroscopic measurements of muonic deuterium and helium transition 
energies as well as additional scattering experiments are expected to shed light on the
problem. For details of these plans and recent works 
we refer the reader to \cite{carlson,antognini}. 
The present article will focus on the theoretical aspects as well as the possible 
discrepancies arising from the methods used for the extraction of the proton radius. 

\section{Proton charge radius and other moments}
The size (or extension) of the proton is characterized by the moments of its charge 
density, $\rho_p$ as 
\begin{equation}\label{newx0}
<r^m> = \int \, r^m \, \rho_p (r) \, d^3 r\, .
\end{equation}
The charge density is conventionally defined as the Fourier 
transform of the electric form factor, $G_E^p({\bm q^2})$, namely,
$G_E^p({\bm q}^2) = \int e^{-i\vec{q} \cdot \vec{r}} \rho_p ({\bf r}) d^3r/(2\pi)^3$. 
Starting with this Fourier transform, 
\begin{eqnarray} \label{newx1}
G_E^p({\bm q}^2) &=&{1 \over 2\pi^2} \int_0^{\infty} r^2 \rho_p(r) 
{\sin(|{\bm q}|r)\over |{\bm q}|r} \, dr \nonumber \\
 &= & {1 \over 2\pi^2} {1\over |{\bm q}|} \, 
\int_0^{\infty} r \rho_p(r) \biggl [ |{\bm q}|r - {|{\bm q}|^3 r^3 \over 6} \, +\, ....
\biggr ]\nonumber \\
&= & {1 \over 2 \pi^2} \, 
\int_0^{\infty} r^2 \rho_p(r) dr - {1 \over 2 \pi^2} \,
{{\bm q}^2 \over 6 } \int_0^{\infty} r^4 \, 
\rho_p(r) dr \, + \, ...\, 
\end{eqnarray}
it is easy to see that the radius 
defined above as $\langle r^2_p \rangle = 
\int \, r^2 \, \rho_p (r) \, d^3 r\, $ can also be 
expressed in terms of the form factor $G_E^p({\bm q}^2)$ as 
\begin{equation} \label{newx2}
 - {6 \over G_E^p(0)} {dG_E^p({\bm q}^2)\over d{\bm q}^2}\biggr|_{{\bm q}^2 =0} =  
\langle r^2_p \rangle.
\end{equation}

There exists another approach in order to extract the proton radius from 
experiment, one involving atomic spectroscopy. 
In this approach, one attempts to calculate the theoretical 
difference between atomic energy levels with the inclusion of 
all possible corrections from quantum electrodynamics (QED) as well as the 
proton finite size corrections (FSC). This difference is then compared with 
the experimentally measured transition energies in order to fit the radius which 
appears in the theoretical expression due to the inclusion of FSC. 
Such an approach was used in \cite{antogpohl} and apart from the second 
moment of the charge density, the FSC in \cite{antogpohl} also
included the third Zemach moment \cite{zemach} defined by 
\begin{equation}\label{thirdZ}
\langle r^3 \rangle_{(2)} = \int \, d^3r \,r^3\, \rho_{(2)}(r)\, ,
\end{equation}
where, $\rho_{(2)}(r) = \int d^3z\,\rho_p(|{\bf z} - {\bf r}|)\, \rho_p(z)$.   
This inclusion introduced a small model dependence in the extraction and in fact
has been discussed at length by several authors \cite{friarsickrujulamiller}.
Some uncertainty depending on the approach to include the FSC was also found 
in \cite{weNPB}. 

\subsection{Breit frame, Lorentz boost and relativistic corrections}

In order to compare the radius extracted from the two methods mentioned in the 
previous subsection, we must ensure that the extractions are done in the same frame 
of reference. As mentioned in \cite{vanderwalcher}, the size and shape of an object
are not relativistically invariant quantities: observers in different frames will 
infer different magnitudes for these quantities. 
The static relation $\langle r^2_p \rangle = 
\int \, r^2 \, \rho_p (r) \, d^3 r\, $, 
defines the radius in the rest frame of the proton. 
The extraction of the radius from electron proton ($ep$) scattering is however 
not done in the proton rest frame.  
The $ep$ scattering data is used to extract the invariant form factor $G_E^p(q^2)$, where 
$q^2 = \omega^2 - {\bf q}^2$ 
is the four-momentum transfer squared in $ep$ elastic scattering. 
The radius is then evaluated using the following relation \cite{bernie}:
\begin{equation} \label{newx3}
\langle r_p^2 \rangle = - {6 \over G_E^p(0)} {dG_E^p({q}^2)\over d{q}^2}
\biggr|_{{q}^2 =0}.
\end{equation}
This definition looks slightly different from that derived in Eq.(\ref{newx2}) 
with the three momentum being replaced by the four momentum squared in (\ref{newx3}). 
At first sight, Eq.(\ref{newx3}) has the appearance of a Lorentz invariant 
quantity (and has even misled some authors to believe so \cite{eidesarxiv}). 
However, if we examine the condition 
$q^2=0$, with $q^2 = \omega^2 - {\bm q}^2$,
it either means that $\omega^2={\bm q}^2 \neq 0$
(in which case we have a real photon) or $\omega=|{\bm q}|=0$.
It is impossible to exchange a real photon in the
t-channel exchange diagram in elastic electron-proton scattering and
hence we have to drop the first possibility. The second
choice involving $\omega = 0$ is, however, 
equivalent to choosing the Breit or the so-called brick-wall 
frame in which the sum of the initial and final proton momentum is zero.
This interpretation is consistent with what we find in the Breit equation 
where the same reference frame has to be chosen.  
The radius extracted in this frame should then be boosted to the proton rest 
frame before comparing it with the one extracted from atomic spectroscopy 
\cite{weJHEP}. This and other relativistic corrections become important \cite{weJHEP} 
with the improved precision of experimental data. Finally, we would like to comment 
that the extraction of the proton radius from atomic spectroscopy relies on formulas 
which start with the definition of the radius as given in (\ref{newx0}).  

The form factor $G_E^p({\bm q}^2)$ is a Fourier transform of the density $\rho_p(r)$ 
in the rest frame and hence $G_E^p({\bm q}^2) = G_E^p(q^2)$ in the non-relativistic 
case but $G_E^p({\bm q}^2) \ne G_E^p(q^2)$ in the relativistic case. 
There have been several attempts in literature in order to incorporate the above 
relations with relativistic corrections \cite{allrel}. 
The fact that the structure of a bound system is independent of its motion in the 
non-relativistic case whereas it changes in the relativistic case depending on 
how fast it moves, was taken into account in \cite{beachey} for the 
calculation of the deuteron radius too. The authors in \cite{weJHEP} found that 
incorporating the relativistic corrections (along with the Lorentz boost) could indeed
remove the 4$\%$ discrepancy between the $e p$ scattering and $\mu p$ Lamb shift 
determinations of the radius. 

\subsection{Finite size effects}
The corrections to the energy levels at order $\alpha^4$ 
due to the structure of the proton are usually included using first order perturbation 
theory with the point-like Coulomb potential modified by the inclusion of form factors 
\cite{weNPB}. 
The determination of the proton radius from accurate Lamb shift measurements 
in \cite{antogpohl} relies for the FSC 
on a seminal calculation of Friar \cite{friarannals} based 
on a third order perturbation expansion of the energy which leads to
an expression which depends on the proton radius rather than the form factors 
explicitly. Such an expression is a result of approximating the atomic wave function 
everywhere by its value at its centre and is useful in extracting 
the radius from spectroscopic measurements. In \cite{friarannals}, the author finds,  
\begin{equation}
\Delta E 
\simeq \langle 0|\Delta V|0\rangle + \langle 0|\Delta V| \Delta \phi \rangle 
+ (\langle \Delta \phi |\Delta V| \Delta \phi \rangle - \langle 0|\Delta V|0\rangle 
\langle \Delta \phi |\Delta \phi \rangle
\end{equation}
where $\Delta V$ is the perturbation and the wave function $|\Psi\rangle = |0\rangle 
+ |\Delta \phi\rangle$ with $|0\rangle$ and $|\Delta \phi \rangle$ the unperturbed part 
and the first order perturbation respectively. Further, approximating the wave function 
$\Phi_n(r) = \langle {\bf r} | 0 \rangle$ by its value at $r = 0$, 
\begin{equation}\label{friarfsc}
\Delta E_{FSC} = {2 \pi \alpha Z \over 3} |\Phi_n(0)|^2\, 
\biggl [\, \langle r^2 \rangle\,  - {\alpha Z m_r \over 2} \langle r^3 \rangle_{(2)} \, 
+ \, ... \biggr ] . 
\end{equation}
The second term involves the third Zemach moment given by Eq.(\ref{thirdZ}) which 
can be rewritten in terms of $\langle r^2_p \rangle$ as: 
\begin{equation}
\langle r^3 \rangle _{(2)} = {48 \over \pi} \int_0^{\infty} {dq\over q^4} \, 
\biggl ( G_E^2(q^2) - 1 + q^2 {\langle r^2_p \rangle \over 3} \biggr )\, .
\end{equation}
The extraction of the radius from the muonic Lamb shift \cite{antogpohl} was done using 
the above relation with a dipole form for $G_E(q^2)$
in order to rewrite $\langle r^3 \rangle_{(2)}$ in 
(\ref{friarfsc}) in terms of $\langle r^2_p \rangle$. Replacing all coefficients 
in (\ref{friarfsc}) and 
including all QED corrections, the final expressions used in the two references 
in \cite{antogpohl} in order to compare with the experimental values 
of $\Delta E (=E_{2S_{1/2}}^{f=1}- E_{2P_{3/2}}^{f=2}) = 206.2949(32)$ meV and
$\Delta E_L (=E_{2P_{1/2}}- E_{2S_{1/2}}) = 202.3706(23)$ meV were

\begin{eqnarray}\label{pohlform}
\Delta E (=E_{2S_{1/2}}^{f=1}- E_{2P_{3/2}}^{f=2}) 
= 209.9779(49) - 5.2262 r_p^2 + 0.0347 r_p^3 \, \, {\rm meV},\nonumber \\
\Delta E_L (=E_{2P_{1/2}}- E_{2S_{1/2}}) = 206.0336(15) - 5.2275(10) r_p^2 + 
\Delta E_{TPE}\, , 
\end{eqnarray}
where the last term corresponds to the full two-photon exchange (TPE) contribution
\cite{antogannals}. Note that the $<r^3>_{(2)}$ term in Friar's expression 
(\ref{friarfsc}) is an order $\alpha^5$ correction and corresponds in principle to 
a two photon exchange diagram as shown in \cite{carlsonvander}. 

In order to confirm that the above formula (\ref{pohlform}) which relies on 
perturbative methods and is 
used to fit the proton 
radius does not change significantly due to the use of nonperturbative methods, 
the authors in \cite{millercheck} calculated  
the transition energies by numerically solving 
the Dirac equation including the finite-size Coulomb interaction and finite-size 
vacuum polarization. 
The point-like Coulomb 
potential was replaced by one including 
the proton charge distribution, $\rho(r)$, given by 
\begin{eqnarray}
V_C(r) = -{Z \alpha \over r} \rightarrow - Z \alpha \int {\rho(r^{\prime})\over 
|\vec{r} - \vec{r}^{\,\prime}|} \, d^3r^{\prime}\nonumber \\
\rho(r) = {\eta \over 8 \pi} \, e^{-\eta r}; \, \, \, \eta = 
\sqrt{{12\over \langle r_p^2\rangle}}\, .
\end{eqnarray}
The energy shift was calculated by taking the difference between 
the eigenvalues calculated using the Dirac equation with the above potential for several 
values of $\langle r_p^2\rangle$. These energy shifts were then interpolated and fitted 
to the function $f = A \langle r_p^2\rangle + B \langle r_p^2\rangle ^{3/2}$, in order 
to determine the coefficients $A$ and $B$. Their final result, namely, 
\begin{equation}\label{mill}
\Delta E (=E_{2S_{1/2}}^{f=1}- E_{2P_{3/2}}^{f=2}) 
= 209.9505 - 5.2345 r_p^2 + 0.0361 r_p^3 \, \, {\rm meV},
\end{equation}
as compared to (\ref{pohlform}) led to a radius which differed from the central value 
of 0.84184(67) fm but was well within the errors. Thus, no significant discrepancy 
between perturbative and nonperturbative methods was found. The authors in \cite{pang}, 
on solving the Schr\"odinger equation numerically however found that the difference  
between perturbative methods and nonperturbative numerical calculations of the 
2$S$ hyperfine splitting in muonic hydrogen are larger than the experimental precision.

A different relativistic approach for the FSC based on the Breit equation with 
form factors was 
investigated in \cite{weNPB}. The method relies on the fact that 
all $\vec{r}$
dependent potentials in Quantum Field Theory (QFT) are obtained
by Fourier transforming an elastic scattering amplitude
suitably expanded in $1/c^2$. The Breit equation \cite{bethe,LLbook,desanctis}
follows the very same principle for elastic
$e^-\mu^+$, $e^+e^-$ (positronium), $e^-p$ (hydrogen) and
$\mu^-p$ (muonic hydrogen) amplitudes.
The one-photon exchange amplitude between
the proton and the muon then leads to the Coulomb potential plus the fine
and hyperfine structure (hfs), the Darwin term and the retarded potentials
\cite{bethe,LLbook}. The authors modified the standard Breit potential 
\cite{weNPB,wejphysg} 
for the $\mu^- p$ system with the inclusion of the 
electromagnetic form factors of the proton. The FSC to the Coulomb, Darwin, fine and 
hyperfine energy levels for any $n,l$ were provided and performing an expansion of the 
atomic wave functions an alternative expression for 
$\Delta E (=E_{2S_{1/2}}^{f=1}- E_{2P_{3/2}}^{f=2})$ was obtained. The main difference 
in their expression as compared to that of \cite{antogpohl} arose due to the inclusion 
of the Darwin term with form factors. Since the use of a Dirac equation for energy 
levels would imply the inclusion of the Darwin term, the authors subtracted the 
point-like Darwin term from their calculations leaving only the effect of this 
relativistic correction with form factors. They obtained 
\begin{equation}\label{ourdelta} 
\Delta E (= E_{2P_{3/2}}^{f=2} - E_{2S_{1/2}}^{f=1})  
= 209.16073 + 0.1174 r_p - 4.2585 r_p^2 + 0.0203 r_p^3 \, \, {\rm meV}, 
\end{equation}
leading to a proton radius of $r_p = 0.83594(46)$ fm which was close to that 
obtained in \cite{antogpohl} but hinted toward an uncertainty introduced due to 
the use of a different FSC approach. 

A brief discussion of the FSC in the hyperfine splitting is in order here. 
The FSC to
the hyperfine splitting in \cite{wejphysg} was evaluated using 
\begin{equation}\label{breit1}
\Delta E_{hfs} = \int |\Phi_C({\bf r})|^2 \,  \hat{V}_{hfs}({\bf r}) \, d{\bf r}
\end{equation}
where $\Phi_C({\bf r})$ is the unperturbed hydrogen atom wave function.
The spin operators are included in the definition of $\hat{V}_{hfs}$ (see 
\cite{wejphysg}). This correction seemed to be different from that used in 
\cite{antogpohl} where it was calculated using the standard 
Zemach formula given by,
\begin{equation}\label{zem1}
\Delta E_{hfs} = - {2\over 3} \mu_1 \, \mu_2 \, <
\mbox{\boldmath$\sigma$}_1
\cdot 
\mbox{\boldmath $\sigma$}_2
> 
\, \int\, |\Phi({\bf r})|^2 \, f_m({\bf r})\, d{\bf r}\, ,
\end{equation}
where $f_m({\bf r})$ is the Fourier transform of $G_M({\bf q}^2)$. 
It was however shown in \cite{wepram} that 
Eqs (\ref{breit1}) and (\ref{zem1}) would give
the same result provided we replace $\Phi_C$ by $\Phi$ in (\ref{breit1}).
Whereas $\Phi_C({\bf r})$ in (\ref{breit1}) is a solution of the 
point-like $1/r$ Coulomb
potential, $\Phi({\bf r})$ is the solution of the potential which includes the 
Coulomb potential with form factors and is given in \cite{zemach} as,
$\Phi({\bf r}) = \Phi_C({\bf r}) + m_1 \alpha \Phi_C(0) \int f_e({\bf u}) 
|{\bf u} - {\bf r} | d{\bf u}$.
The difference thus lies in the usage of
the unperturbed wave function in the energy correction. 
In other words, in \cite{wejphysg,wepram}, the total Hamiltonian is taken as
$H = H_0 \, + \, H_C^{FF} \, +\, H_{hfs}^{FF}$
with $H_0$ containing the $1/r$ Coulomb potential, $H_C^{FF}$, the finite size correction
to the Coulomb potential and $H_{hfs}^{FF}$ the hyperfine interaction with form factors
leading to the energy correction in first order perturbation theory given by 
$\Delta E = <\Phi_C|H_C^{FF} | \Phi_C> \, +\, <\Phi_C |H_{hfs}^{FF} | \Phi_C>$.
In \cite{zemach} however, one finds $H = \tilde{H}_0 +\, H_{hfs}^{FF}$, with
$\tilde{H}_0$ which includes FSC to the Coulomb potential
taken as the unperturbed Hamiltonian.
We notice from the above discussion that the Breit equation and the Zemach method would
lead to the same hyperfine correction if the time independent perturbation theory is
handled in the same way. 
In a calculation which involves finite size corrections to 
the point-like Coulomb potential as well as hyperfine structure
taken separately (as in \cite{wejphysg,antogpohl})
it seems reasonable to use the prescription with
$\Delta E = <\Phi_C|H_C^{FF} | \Phi_C> \, +\, <\Phi_C |H_{hfs}^{FF} | \Phi_C>$
in order to avoid double counting of the finite size corrections to the Coulomb term.
The $r_p^2$ and $r_p^3$ terms in Eqs (\ref{ourdelta}, \ref{pohlform})
for example appear after the explicit inclusion
of the FSC to the (1/r) Coulomb potential.

The proton radius extracted from the muonic hydrogen Lamb shift is much more accurate 
than that determined from standard (electronic) hydrogen. The procedure of extracting 
the radius from electronic hydrogen is slightly different and involves a simultaneous 
determination of the Rydberg constant and Lamb shift. Traditionally, the Lamb shift 
was actually a splitting (and not a shift) between the energy levels $E(2S_{1/2})$ and 
$E(2P_{1/2})$ which are degenerate according to the naive Dirac equation in the 
Coulomb field. The convention now is however to define the Lamb shift as any deviation 
from the prediction of the naive Dirac equation 
that arises from radiative, recoil, nuclear structure, relativistic and binding 
effects (excluding hyperfine contributions) \cite{eidesreport} so that, 
$E_{njl} = E^{Dirac}_{nj} + L_{njl}$. The measurement of the Lamb shift can be 
disentangled from the Rydberg constant by using two different intervals of 
hydrogen structure. For example, we can use the accurate measurements of 
$f_{1S-2S} = 2466061413187.34(84)$ kHz and 
$f_{2S_{1/2} - 8D_{5/2}} = 770649561581.1(5.9)$ kHz along with the energy expressions 
\begin{eqnarray}
E_{1S-2S} = [ E^{Dirac}_{2S_{1/2}} - E^{Dirac}_{1S_{1/2}}] + L_{2S_{1/2}} - 
L_{1S_{1/2}} \nonumber \\
E_{2S-8D} = [E^{Dirac}_{8D_{5/2}} - E^{Dirac}_{2S_{1/2}}] + L_{8D_{5/2}} - 
L_{2S_{1/2}}\, ,
\end{eqnarray}
to determine the radius. The first differences on the right hand side are dependent on 
the Rydberg constant $R_{\infty}$ (through $E^{Dirac}_{nj} = R_{\infty} E_{nj}$), 
which can be eliminated using the two equations. The left hand 
side is replaced by accurate measurements and the Lamb shift is determined independent 
of the Rydberg constant. Knowing the accurate value of the Lamb shift, it can be 
inserted back into the above equations to determine the Rydberg constant accurately. 
The value of the Rydberg constant is thus obtained to be \cite{codata} 
$R_{\infty} = 10973731.568539(55)$ m$^{-1}$. Knowing $R_{\infty}$ accurately, one can 
now proceed to determine the radius as follows:
$$ {\rm Measured \, \, energy \, \, splitting} = R_{\infty} \, E_{nj} \, + 
\, E({\rm Lamb \, \,shift})$$
where $E({\rm Lamb \, \,shift})$ includes all QED as well as 
proton structure corrections. With a good knowledge of all QED related corrections 
(see for example \cite{martynall}), the radius in the proton structure corrections 
appearing in $E({\rm Lamb \, \,shift})$ can be fitted to the measured energy splitting. 

\section{Reanalyses of scattering data} 

Apart from the various theoretical papers which attempted to explain the discrepancy 
between the proton radius from spectroscopy and scattering, there have also been 
some attempts at reanalysing the electron proton scattering data. Here, we shall mention 
some of the recent works and the criticisms too. In \cite{horbat}, the cross sections 
at the lowest $q^2$ were fitted using two single parameter models for form factors 
with one being the standard dipole given by $G_E^2(q^2) = (1 + q^2/b_E)^{-4}$, 
$G_M^2(q^2)/\mu_p^2 = (1 + q^2/b_M)^{-4}$ and the other involving a 
Taylor expansion given as, $G_E^2 = 1 - c_E z$, $G_M^2/\mu_p^2 =  1 - c_M z$, where 
$z$ is the conformal mapping variable as defined in \cite{horbat}. Following the 
philosophy that the charge radius of the proton is a small-$q^2$ concept, the authors 
analysed the low $q^2$ data using the simple fits and reached the conclusion that 
the proton radius could vary between 0.84 and 0.89 fm, thus making the spectroscopy 
and scattering results consistent. 

In another attempt of a similar kind, instead of focussing on a reanalysis of 
recent data, the authors decided to review the older Mainz and Saskatoon data 
in \cite{higin}. They found that a dipole function with the muonic hydrogen radius 
of 0.84 fm, i.e., $G_E(q^2) = (1 + q^2 /0.66 [{\rm GeV ^2}])^{-2}$, not only describes 
low $q^2$ $G_E(q^2)$ results, but also reasonably describes $G_E(q^2)$ to the highest 
measured $q^2$. 
The authors in 
\cite{higin,grifi} performed a sharp truncation of the form factor expansion in momentum 
space which was strongly 
criticized for not being in accord with the basic facts of form factors 
and the extraction of radii from them in \cite{distlerwalcher}. 

A completely novel point of view was chosen in \cite{terry} where the authors 
noted that the proton radius may not be unique but a quantity which is randomly 
distributed over a certain range. The standard definition of a ``radius" of the proton 
is obviously based on the notion of the proton being spherical. Arguing that the 
definition of the radius could get blurred for a deformed proton and providing 
other literature in support of the idea of a fluctuating size of the proton, the 
authors performed a fit for a form factor of the so-called ``non-identical"
protons. Taking the standard dipole form factor as the basis, the authors introduced 
the fluctuation of the proton size by performing an average with the following form:
\begin{equation}\label{terryG}
\langle G_E^2(q^2,\Lambda_1)\rangle = {1 \over 2 \Delta \Lambda} 
\, \int_{\Lambda_1 - \Delta \Lambda} ^{\Lambda_1  + \Delta \Lambda} \, 
G_E^2(q^2, \Lambda) d\Lambda\, ,
\end{equation}
with the $G_E$ in the integrand having the standard dipole form. 
Using the latest Mainz data to perform the fits, the authors determined an average 
$\Lambda = 0.8203$ GeV with a variation $\Delta \Lambda$ of 21.5\%. They further 
studied the effects of such a radius variation in neutron star and symmetric nuclear 
matter. The electric form factor as defined in (\ref{terryG}) can be evaluated 
analytically and using Eq. (\ref{newx3}) leads to a radius given by 
\begin{equation}
r_p^2 = {12 \over \Lambda_1^2 - \Delta\Lambda^2}\, , 
\end{equation}
which with the substitution of the values from \cite{terry} gives a proton radius, 
$r_p = 0.864$ fm. On applying the relativistic correction 
(involving the Lorentz boost with $\lambda_E = 1$) 
in \cite{weJHEP}, the radius reduces to a value of 0.844 fm which is quite close to 
that determined from muonic hydrogen spectroscopy \cite{antogpohl}.  
In Fig. 1 we display the proton electric form factor at low momenta within the 
the three different parametrizations discussed above. In \cite{terry}, the authors 
investigated the density dependence of the proton radius in nuclear matter.
The right panel in the figure shows the behaviour of the proton radius 
using the parametrization in \cite{terry}, with and 
without relativistic corrections (as found in \cite{weJHEP}). 

\begin{figure}
\begin{center}
\includegraphics[width=17cm,height=7cm]{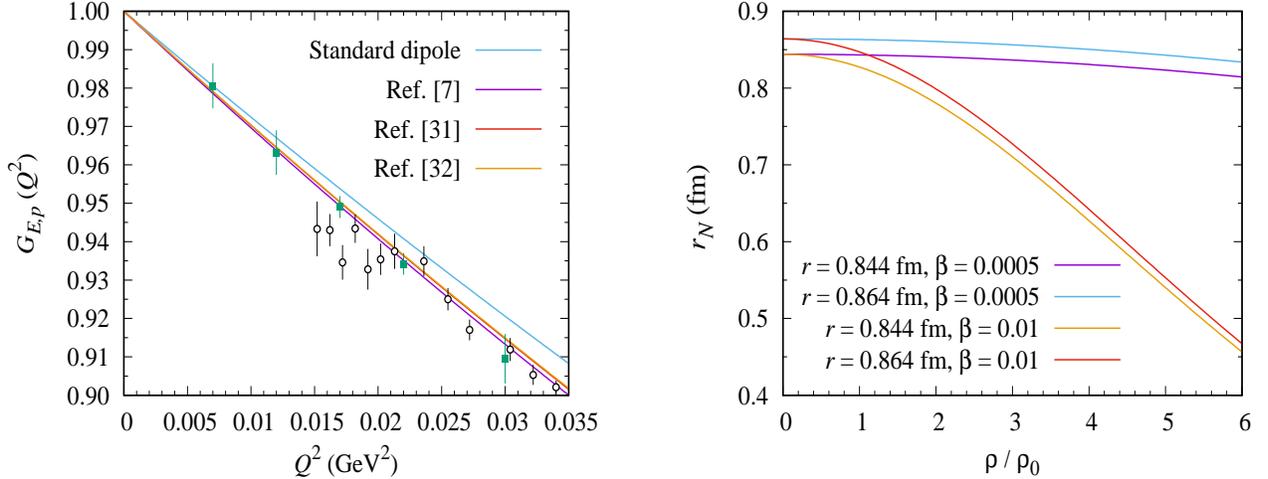}
\caption{Comparison of the parametrizations 
from Refs \cite{horbat,higin,terry} for form factors at low $q^2$ (shown in the 
left panel). The right side panel displays the density dependent proton radius
as calculated in \cite{terry} but with and without relativistic corrections included.}
\label{fig:bimg}
\end{center}
\end{figure}

\section{Brief overview of planned experiments}
The discussion of the proton radius puzzle has so far revolved around the 
extractions from $ep$ scattering measurements, standard hydrogen (electronic) 
as well as muonic hydrogen spectroscopy. The missing component in these analyses is then 
the data on muon-proton elastic scattering. The MUon proton Scattering Experiment 
(MUSE) at the Paul Scherrer Institute is a simultaneous measurement of the $\mu^+ p$ 
and $e^+ p$ elastic scattering. The experiment is expected to decide if the $\mu p$ 
scattering and $\mu p$ Lamb shift experiment lead to the same proton radius. 
Another scattering experiment is the PRad which will measure the $ep$ scattering 
cross sections with higher precision and at low $q^2$. Besides these plans, the 
CREMA collaboration has been studying the spectroscopy of other exotic atoms such 
as muonic deuterium and muonic helium too. A detailed account of the future 
experiments can be found in \cite{carlson,antognini}.

\section{Summary}
The finite size of the proton is characterized fully by all the moments of its charge 
distribution. The second moment is however generally used to define the ``radius" of 
the proton. The radius thus defined can either be extracted from spectroscopic 
measurements or lepton proton scattering data using theoretical methods. 
Until some time ago, there seemed to be an agreement between the radii 
extracted from spectroscopy (with standard electronic hydrogen) and scattering. 
However, the high precision muonic hydrogen spectroscopy revealed a 4\% 
deviation from the average value obtained from all previous experiments. 
Since the radius is an extracted and not directly measured quantity, 
a higher experimental precision should also be complemented by a higher confidence 
in the theoretical component. With this viewpoint, in this review we have examined the 
theoretical methods used for the extraction of the radius as well as the related 
literature which appeared in the form of possible solutions of the 
``proton radius puzzle". These included checks on the validity of the perturbative 
methods used, approximations therein and the relevance of relativistic corrections. 
The latter is particularly of importance due to the fact that the relation between 
the charge density and the electric form factor is necessarily of a non-relativistic 
nature. This fact also makes it important that the comparison of radii extracted 
from different experiments is done in the same frame of reference. 
While the resolution of the puzzle is being attempted by reanalyses of old data and 
planning of new experiments, it is necessary to pay attention to the theoretical inputs 
involved in the extraction of the radius too.
\begin{acknowledgments}
T.M. was supported by the Research-Cluster-Grant-Program of the University of Indonesia, 
under Contract No. 1862/UN.R12/HKP.05.00/2015.
\end{acknowledgments}

\end{document}